\documentclass{article}
\usepackage[utf8]{inputenc}
\usepackage{graphicx}
\usepackage{booktabs}
\usepackage{enumerate}
\usepackage[english]{babel}
\usepackage{csquotes}
\usepackage[
backend=biber,
style=numeric,
]{biblatex}
\bibliography{dpo}

\title{We Are Not There Yet: The Implications of Insufficient Knowledge Management for Organisational Compliance}
\author{%
  Thomas Şerban von Davier; Konrad Kollnig; Reuben Binns;\\ Max Van Kleek; Nigel Shadbolt \\
  Department of Computer Science, University of Oxford\\
  Oxford, United Kingdom \\
  Contact: \texttt{thomas.von.davier@cs.ox.ac.uk} \\
}
\date{January 2023}

\begin{document}

\maketitle

\begin{abstract}
Since GDPR went into effect in 2018, many other data protection and privacy regulations have been released. With the new regulation, there has been an associated increase in industry professionals focused on data protection and privacy.  Building on related work showing the potential benefits of knowledge management in organisational compliance and privacy engineering, this paper presents the findings of an exploratory qualitative study with data protection officers and other privacy professionals. We found issues with knowledge management to be the underlying challenge of our participants’ feedback. Our participants noted four categories of feedback: (1) a perceived disconnect between regulation and practice; (2) a general lack of clear job description; (3) the need for data protection and privacy to be involved at every level of an organisation; (4) knowledge management tools exist but are not used effectively. This paper questions what knowledge management or automation solutions may prove to be effective in establishing better computer-supported work environments.
\end{abstract}

\section{Introduction}
When GDPR first went into effect in 2018, Sirur et al. asked, “Are we there yet?” regarding the ability of the industry to comply with the new legislation \autocite{Sirur2018AreGDPR}. They found limited concerns within the private sector about ensuring compliance in larger, well-equipped organisations, with additional concerns in smaller organisations \autocite{Sirur2018AreGDPR}. However, the landscape of recent laws regarding data privacy and online safety has expanded rapidly to include GDPR, Brazil's LGPD, California's CCPA, the planned EU AI Act, and various proposed online harms legislation. As a result, the tech industry must keep track of and comply with a significant body of laws, which motivated us to explore the experiences of relevant industry experts.

% While the technology sector has followed Moore’s Law closely in accelerating its growth at breathtaking speeds, the role of regulation, policy, and the rule of law has been far slower. Part of this discrepancy has been the need for policy and regulation to be reactive to avoid over-reaching and encroaching on the wishes of the general public without any justification \autocite{Kubat2019BalancingIndustries}. This has led to limited opportunities for regulation to be proactive and move ahead of the rapidly accelerating technology growth of the past few decades resulting in a communication and knowledge gap \autocite{Azhar2021ExponentialIt}.

Attempts within organisations to consider privacy and data protection often have engineers and designers collaborate with lawyers and policy experts. However, studies going back as far as the early 2000s have shown that the teams building technology do not feel as though privacy and data protection are among their primary concerns and obligations \autocite{Langheinrich2002D15.3.1-AWP15-Privacy,Bednar2019EngineeringChallenge}. As a response, researchers suggested a range of value-based software engineering frameworks\autocite{Spiekermann2009EngineeringPrivacy,Shilton2013ValuesDesign,Spiekermann2020Value-basedDesign}. As privacy data protection laws have created new roles within organisations, notably data protection officers (DPOs) and other privacy professionals, we see the opportunity to revisit some of these past discussions, especially in light of literature exploring the benefits of knowledge management (KM) on compliance behaviour \autocite{Gressgard2014KnowledgeSystem, Boella2013ManagingCompliance, Wipawayangkool2009InformationDiversification, Kim2017TheBehavior}.
% regarding the role of regulatory privacy compliance within the tech industry.

DPOs and privacy professionals within organisations are the primary parties on the receiving end of regulations crafted by legislators. Their perspective on implementing applicable privacy laws is invaluable for crafting future regulations and ameliorating challenges and weaknesses in the current system. Furthermore, their organisations and external parties often consider these industry professionals accountable for ensuring regulatory compliance. This motivates us, in this piece, to identify what role KM plays in the DPO role and where there might be room for improvements. In our work, we focus on the three, related research questions:
\begin{enumerate}[\textbf{{RQ}:}]
    % \item \textbf{RQ1:} How do industry experts in data protection and privacy bridge the regulation-practice divide to ensure regulatory compliance is met within their organisations?
    \item (Knowledge) How do data privacy experts bridge multiple knowledge bases from various teams across an organisation?
    % \item \textbf{RQ2:} How do data protection officers maintain and communicate the knowledge needed for regulatory compliance within an organisation?
    \item (Communication) How do data protection officers maintain and communicate compliance decisions made internally and externally to the organisation?
    % \item \textbf{RQ3:} How do data protection officers foresee changes to their role over time?
    \item (Experiences) What are the current successes and failures data protection officers experience while working to ensure compliance?
\end{enumerate}

Our preliminary discussions with experts in the industry that fill data protection (DPOs), cybersecurity, or legal roles revealed that they often assume the role of auditors and critics within their organisation. They are responsible for improving the internal practices to ensure that any data breach or other emergency scenario is handled effectively and within the standards outlined by law. They also need to occasionally oppose internal decisions where they breach relevant law.

Ultimately, our interviews revealed four categories of challenges associated with KM and the role of privacy professionals. At a high level, these categories are: (1) a perceived disconnect between how businesses operate and how regulation is implemented; (2) a lack of a clear job description; (3) a need for data protection and privacy to be involved and communicated at every level of an organisation; (4) a common availability but ineffective use of knowledge management tools. By establishing our work within the current canon of related research in organisational compliance and privacy engineering attitudes, we will present the perspectives of our participants, revealing the need for advances in knowledge management. We will then discuss how future work in computer-aided knowledge management can improve our lives through added compliance with privacy regulations.

\section{Related Work}

There are two predominant areas of research that serve to contextualise and motivate our research. The first is existing research on organisational compliance with regulatory bodies. The second is a body of work that has explored privacy engineering attitudes with an established set of methods.

\subsection{The Role of Knowledge Management in Organisational Compliance}

In response to innovation, society has started developing rules and regulations to minimise potential dangers. The history of data protection and privacy is closely tied to advances in technological innovation \autocite{Holvast2007HistoryPrivacy}. As our technological capabilities and data collection behaviours increase, the need for careful, privacy-preserving legislation also increases. These regulations often include detailed legal documents and associated lists of standards that need to be processed, understood, and implemented. Furthermore, ensuring compliance with these regulations is an interdisciplinary effort across an organisation.

In many ways, the need for compliance and regulation is nothing new. Specific fields like pharmaceuticals and hazardous industries have been controlled and regulated almost continuously \autocite{Kubat2019BalancingIndustries, Hopkins2011Risk-managementIndustries, Papaluca2013GatekeepersAttitude, Drimmer2010HumanTrends}. Naturally, there is an observed, unofficial correlation between the strength of regulation and the danger posed by the field being regulated. Similarly, we see a demonstrated history of regulation in place for the financial market \autocite{Crockett2000Commentary:Integration, Appaya2020HowBeyond}. These fields and the associated papers demonstrate a history of organisational compliance analysis. Furthermore, these fields contain a notable body of research into the effectiveness of KM on compliance across an organisation's employees \autocite{Gressgard2014KnowledgeSystem, Boella2013ManagingCompliance, Wipawayangkool2009InformationDiversification}. The findings indicate the potential for more effective KM to be correlated with improved compliance behaviours. Through rigorous study and review of the fields, we can better understand the impact of regulation and continue to steer the system away from ineffective solutions.

A notable portion of the research done on organisational compliance focuses on what regulators need to do or how they should respond to certain changes in the field \autocite{Papaluca2013GatekeepersAttitude, Crockett2000Commentary:Integration, Appaya2020HowBeyond, Drimmer2010HumanTrends}. For example, some regulatory theory argues that regulators need to be ``really responsive" in their process by taking into account the actions, frameworks, and attitudes of the organisations they are regulating \autocite{Baldwin2008ReallyRegulation}. While it is undoubtedly valuable to elevate items of concern to the attention of regulators, there is often limited analysis of what steps an organisation can take to improve their regulatory compliance \autocite{Hopkins2011Risk-managementIndustries, Kubat2019BalancingIndustries}. These analyses often highlight the challenges organisations face by balancing compliance with expansion and improvement and the risks involved with these decisions. Some articles %from the Harvard Business Review
take a step further by implying that the decision to comply with regulations is relatively light and can be easily dismissed \autocite{Cannon2014HowRegulators, Chen2018WhyThem}. One such article advocates for “winning over regulators” rather than working to understand the need for regulation in the first place. Another highlights the danger of compliance becoming a passive “checklist” activity. To solve the potential issue of compliance becoming a box-ticking exercise, regulatory theory argues that through discourse between organisations and legislators compliance and legitimacy can become a more interactive process and as a result, a more effective process \autocite{Black2002RegulatoryConversations, MasonBurdon2020InstitutionalSector}. In light of this research at the organisational level, our research intends to explore the specific actions taken by professionals responsible for regulatory compliance within an organisation, especially in data protection and privacy.

Some work within the area of data protection and privacy in conjunction with industry professionals has been done within the context of energy company employees and drone operators and manufacturers  \autocite{Finn2016PrivacyOrganisations, Clarke2014ThePrivacy}. Here there is further support of the correlated impact between KM and compliance behaviour\autocite{Kim2017TheBehavior}. These research initiatives explored how KM can be used across a broad range of employees in their individual compliance tasks. Our research aims to explore the role of KM for a specific subset of data regulation and privacy specialists.

\subsection{Value-Based Privacy Engineering}

While data protection and privacy have grown into regulation and legal discussions, a body of research has already been done to improve the considerations of privacy in design and engineering. An early study by Lahlou and Langheinrich opened the door to exploring how engineers view their role in ensuring data protection and privacy \autocite{Langheinrich2002D15.3.1-AWP15-Privacy, SaadiLahlou2005PrivacyComputers}. To clarify, the general term “engineers” used in this paper and subsequent work refers to all technical product team members involved in building a system, such as designers, systems engineers, and requirement engineers. This early work led to a set of methods and approaches to design and engineering that continue to generate discourse in the field of ethical computing \autocite{Lahlou2004EuropeanAgoras, Spiekermann2009EngineeringPrivacy, Nissenbaum2010PrivacyLife}. However, comparing and analysing the effectiveness and usage of these frameworks lie outside this paper's scope.

These frameworks have been explored and tested via a selection of methods aimed at understanding the attitudes and behaviours of industry professionals concerning data protection and privacy. We reviewed a collection of these methods to contextualise our approach within this canon of research. Shilton used a participant-observer method within an academic design and research team to explore what experiences and interactions were effective in introducing ethics into the design process \autocite{Shilton2013ValuesDesign}. Their combination of fly-on-the-wall observations with follow-up interviews provided fascinating insights. Nonetheless, their participants were in an open academic environment, whereas our interested population is industry professionals dealing with proprietary information. Some research on industry professionals has been done by reviewing anonymous posts on a professional discussion forum investigating privacy practices of mobile developers \autocite{Shilton2017LinkingDevelopment}. The challenge is that, unlike organised iOS or Android forums, there are limited forums for privacy professionals within the tech sector for us to use for our research.

Ultimately it was recent work that combined survey data with interview data that provided us with a potential method of investigation. To better understand the challenges experienced by engineers, there was a two-part project conducted on professionals from Ubicomp as well as select senior-level engineers \autocite{Spiekermann2019InsideEngineers, Bednar2019EngineeringChallenge}. This mixed-method approach effectively revealed attitudes towards privacy and responsibility felt by many engineers. We decided to take a similar approach with our methodology with some adjustments to fit our particular research interests.

From this research into potential methods regarding industry professionals, we were motivated to focus on a slightly different population that has grown since GDPR came into force in 2018. This population of data protection officers are part of the technical teams that have been explored in previous work, but their roles are officially associated with regulatory compliance. Again, we can look to previous research done at the outset of GDPR \autocite{Sirur2018AreGDPR}. They interview industry professionals before the establishment of DPOs. Their findings were hesitantly optimistic, citing that overall, organisations felt comfortable with regulatory compliance expectations, with some concerns among smaller organisations not working in the tech or security sectors \autocite{Sirur2018AreGDPR}. Since the 2018 paper, there remains a need for further investigation into how DPOs are functioning in their roles within organisations. A 2021 survey in Croatia found 82\% of their sampled DPOs reported they did not have enough education, knowledge, or training needed to feel ready for their role \autocite{Mladinic2021Post-gdprStudy}. Thus our interview study aims to reveal specific discussions regarding attitudes and approaches to organisational data protection and privacy.

\section{Methods}
This outlines the details involved in creating and executing our qualitative exploratory study. We were inspired by the methods Spiekermann et al. and Bednar used in related work to elicit attitudes and thoughts directly from the participants \autocite{Spiekermann2019InsideEngineers, Bednar2019EngineeringChallenge, Sirur2018AreGDPR}. The primary stakeholders we were interested in meeting with were data protection officers, project managers, and any developers or legal team members working with data regulation compliance.

\subsection{Recruitment}
To gather our participants, we utilised professional online networks that would reach industry professionals working within organisations across the USA and UK. We recruited through direct emails, messages, and a larger post calling for participants shared on LinkedIn and Twitter. We worked to provide the participants with the option to either discuss online in real-time or fill out the interview questions as an online form.
 
We got insights from a total of eight participants, which builds on the numbers of previous work \autocite{Langheinrich2002D15.3.1-AWP15-Privacy, Bednar2019EngineeringChallenge}. Six participants were interviewed, with the remaining two filling out the interview questions via the form. The participants represented a range of data protection and privacy professionals from early career to C-suite (see Table \ref{tab:experts}). Their companies ranged from small startup sizes to large international organisations.

\begin{table}[]
    \centering
    \caption{This outlines the IDs and Gender of our participants, a short description of their current role, and their primary location of operation.}
    \begin{tabular}{llll}
        \toprule
        ID (Gender) & Role & Location & Company Size \\
        \midrule
        P1 (M) & Data protection officer & USA & 250-500 \\
        P2 (M) & Regulatory leader & UK & 1000+ \\
        P3 (F) & Legal and security operator & UK & 50-100 \\
        P4 (M) & Founder of software company & UK & 50-100 \\
        P5 (F) & Associate data protection officer & EU & 1000+ \\
        P6 (M) & Project manager & USA & 1000+ \\
        P7 (unknown) & Legal and privacy professional & USA & 250-500 \\
        P8 (unknown) & VP of privacy & USA & 1000+ \\
        \bottomrule
    \end{tabular}
    \label{tab:experts}
\end{table}

\subsection{Procedure}
The primary method was to conduct a virtual interview between one of our researchers and an industry professional. The interview was designed to understand how the professionals approach regulatory compliance within their role as specialised professionals (Appendix \ref{appendix}). Under a semi-structured interview, we could ask follow-up questions for deeper understanding. The second was an online form meant as the “lighter” of the two ways to participate in our research (Appendix \ref{appendix}). It was essentially identical to the interview but without the possibility of a follow-up. It also provided the participants with additional anonymity.
 
We spent roughly 30--40 minutes with each participant in the interview. Using Microsoft Teams for the online interviews allowed us to use the automatic transcription tool. This avoided the need to capture any permanent audio or video recordings of the participant to ensure their anonymity. Furthermore, the transcripts were further anonymised to remove any reference to organisation or participant names. Our work to ensure and maintain anonymity is one of the reasons we successfully accessed individuals in privileged positions within their organisations. Section \ref{policy} considers why our anonymity commitment was vital for our participants. Additionally, we insisted that we were not interested in proprietary organisational information but rather in the anecdotes and learnings of the participants themselves. They were the experts dealing with internal and external policies, and we wanted to hear from them.

\subsection{Data Analysis}
Following the collection of information from the participants, we conducted an affinity clustering exercise similar to that found in other HCI methods \autocite{Holtzblatt2007ContextualDesign}. Affinity clustering involves breaking down the interview transcripts into single ideas or statements. All of the individual statements of the participants are then grouped by similarity. Each group is labelled, creating a new layer, and the process is repeated by grouping the groups until one or two coherent theses form. We reached a high degree of data saturation at around the sixth participant. In other words, we started hearing the same feedback and perspective multiple times, indicating shared experiences.

Applying affinity clustering to the transcripts, we were able to identify three layers of themes regarding the practice of ensuring compliance with regulatory bodies. This method allows us to consolidate feedback from multiple participants under a single idea which then gets grouped with other ideas to form a broader theme or message. Looking across all the themes, we could generate an overall thesis of our findings.

% \begin{figure}
%     \centering
%     \includegraphics[width=9cm]{Images/DPO_Affinity.png}
%     \caption{The above figure is an image of the interview findings organized in an HCI affinity diagram. \protect}
%     \label{fig: viz0}
% \end{figure}

\section{Findings}
The feedback from the participants regarding their experiences highlighted four major categories of challenges. Within each major category, we recorded their thoughts and drew them together and highlighted how advances in knowledge management can address the experiences of our participants.  

% \begin{figure}
%     \centering
%     \includegraphics[width=9cm]{Images/DPO_Visual1.png}
%     \caption{Comments regarding industry professionals' views regarding some recent regulation. \protect}
%     \label{fig: viz1}
% \end{figure}

\subsection{Regulation Disconnect}
Many industry professionals mentioned feeling a lack of understanding from regulators regarding the approach to real-world implementation of the policies. Our participants stated that implementing specific regulations was easier than others. For example, our participants noted how dealing with cookies is one of the most difficult aspects of their job. Our participants stated,

\begin{quote}
    P3: The laws are very nicely written. But if you are operating a business, how do you put that writing into practice? There's a lot of grey area I mean. Cookies is one great example.
\end{quote}

\begin{quote}
    P4: I would say one example is the cookies … situation about cookies because it's a bit of a mess. The rules are not fully clear. They've changed over time. 
\end{quote}
On the other hand, the same participant mentioned how easy it is to establish proper privacy notices by saying,
\begin{quote}
    P3: The processing is very straightforward for privacy notices.
\end{quote}
These selections serve as examples of the differences in implementing data privacy regulation. Both privacy notices and cookie management are common aspects of digital experiences, yet our participants insist there is a difference in how easy they are to implement. While there was some frustration about having different experiences with different types of regulation, there remained a belief that understanding the regulation was important for others in the business. The professionals we met recognised the value of performing consistent regulatory practices.
\begin{quote}
    P3: I think the key for DPIAs is to really get everyone on track. You know it shouldn't be just a tick. I think when it comes to product development. All the people need to really understand data impact assessments.
\end{quote}
Data Protection Impact Assessments (DPIAs) were a common talking point during the interviews as they served as both an essential piece of regulatory compliance and a significant job responsibility for our participants. DPIAs, as stated by the participant, required involvement from multiple teams across the organisation, like cookies and privacy notices. Nonetheless, some standard regulatory practices posed greater challenges than others.
 
The difference in effort needed to achieve certain regulatory practices was further complicated by slight differences introduced by laws in different nations and regions. The challenge with the current regulation of the Internet is the fact that cyberspace often crosses international borders. One participant said,
\begin{quote}
    P4: I do think that the transfer impact stuff is a little bit onerous because the European Union essentially said that you have to make your own judgment on the laws of a foreign country, and that is difficult. For example, if you want to send data to India, then you have to make your own judgment about Indian law, which seems a bit silly that thousands or millions of people should do that
\end{quote}
This challenge of laws and regulations altering from one nation to another can be particularly difficult for smaller companies that might not have international regulation experts on their payroll.
\begin{quote}
    P3: There needs to be a bit more help on these things especially for small organizations I would say.
\end{quote}
A concern for small businesses reflects early concerns presented by related research at the start of GDPR \autocite{Sirur2018AreGDPR}. Considering the feedback has not changed, there may be a significant unsolved problem for small organisations and ensuring compliance. While there exists ambiguity and assumptions as policies are translated from theory into practice, our participants noted they are open to working further with regulators to better establish this process from law to practice. Again, this mirrors previous interview research that identified the soft and hard power government entities have on the practice of privacy and data protection \autocite{Bednar2019EngineeringChallenge}. Our participants explained why they would want to collaborate with policymakers.
\begin{quote}
    P2: So I think engaging with these kinds of bodies like standards bodies is also one of the ways to both understand the space and contribute to helping with the how these requirements are being shaped.
\end{quote}
The privacy professionals were interested in furthering their understanding and contributing their experience to future changes. However, the path to collaboration is not always easy for privacy professionals. One participant mentioned that they could not interact directly with any regulators through their position as only the legal team from their company handles those conversations. In contrast, other participants noted that they often engage with academics and policymakers as part of their role.
\begin{quote}
    P2: As part of this work with a university we started engaging with parliamentary committees. Responding to inquiries and consultations. Including for instance, when the Investigatory Powers Bill came out. It was being discussed, providing some input on that and getting invited to have interview with the Home Office for them to get a better view as to what the potential concerns might be.
\end{quote}
These contrasting views on interacting with policymakers pointed at a larger difference between our participants, how they define their role and how they function within their organisations.

% \begin{figure}
%     \centering
%     \includegraphics[width=9cm]{Images/DPO_Visual2.png}
%     \caption{Comments regarding industry professionals' desire to help with future regulation. \protect}
%     \label{fig: viz2}
% \end{figure}

\subsection{Defining the Job}
One set of findings from our interviews revealed the vast flexibility and resulting uncertainty that comes with the role of a DPO or privacy professional. While governments and legal forces define the focus of DPO work, the actual methods, tools, and job description is still defined by the individual organisations. An example provided by the participants was concerning the process of conducting proper audits.
\begin{quote}
    P2: There is a lack of clarity with the methodology that will be expected. We've seen this in a number of cases, you know with the, in New York with the auditing of uh recruitment algorithms. There's no actual standard for that kind of audit.
\end{quote}
They mentioned that the various guidelines imposed by legislation from different regions seem too broad that everyone is simply conducting audits as they see fit or as their organisation believes is correct. This finding parallels previous work in privacy engineering, where certain engineers identified the problem of “operationalising privacy” \autocite{Bednar2019EngineeringChallenge}. This work highlighted that regulatory forces establish the values and rules regarding privacy outside the organisation, but the actual implementation and methods are predominantly up to the employees. One participant said their strategy was to combine data protection and privacy methods with those of cybersecurity. This is one approach to solving the uncertainty of conducting a proper data audit.
\begin{quote}
    P4: We review and audit internally we are also combined with our security.
\end{quote}
Furthermore, the broader role of privacy professionals within the company can vary quite drastically depending on its position and its business model. Participants provide two contrasting examples.
\begin{quote}
    P3: I remember I was in a conference and there was a big discussion on transfer impact assessments. I mean, how far do you need to go? You know if you a small business, how far would you go to look at all your subprocesses and you know it's just impossible in a sense.
\end{quote}
This participant was sharing how the depth and scale of the work for an impact assessment may be quite different for a privacy professional within a smaller company compared to a larger one with more resources.
\begin{quote}
    P2: During our work there is the need to make sure that you're not going to take a very strong position say against the business model of a major kind of client.
\end{quote}
Here we see the impact the business model of a large, publicly traded company can have on the role of a privacy professional that is attempting to implement regulatory compliance. Between these two pieces of feedback, we see a difference in the agency offered to the professionals. In a smaller company, the DPO might have greater control in implementing a philosophy and policy of data protection and management but suffer a trade-off in terms of resources and overall security of the data being handled. On the other hand, a DPO within a large corporation may not have to worry too much about security and resources but is instead just one cog in a massive shareholder-driven machine. The explicit role of a DPO or privacy professional within a company appears to be flexible and heavily defined by the organisation with which they work and the roles taken by their colleagues in regulatory practices.

\subsection{All in This Together}
All of our interview participants shared that they worked on multiple projects and initiatives across their organisation. Therefore, the work of ensuring regulatory compliance for data protection and privacy involved clear communication across multiple teams. A participant described one approach:
\begin{quote}
    P4: I mean definitely you need different language and ways of presenting things with different audiences.
\end{quote}
Most participants' primary audiences were the development, executive, legal, and marketing teams. One participant gave an example of how language needs to be used for the development team,
\begin{quote}
    P3: I wouldn't go and say ohh I need this data impact assessment to my developers. You know, asking them, “What's your legal basis for processing” I mean I really need to simplify things in their language if you wanna them to answer you back in correct ways.
\end{quote}
Similarly, executives often need some form of simplification. One participant described these types of communications as “very consumable nuggets”. Regardless of the team involved, the DPOs and privacy professionals interact with the entire organisation.
 
Due to this almost universal interaction, the effectiveness of the DPO’s policy implementations is also dependent on the reception of others within an organisation. In some cases, data protection and privacy are established parts of the company culture.
\begin{quote}
    P4: So we have a we have a very strong internal data protection policy. We have a formal policy. We train on it. Everybody has to take a test on it every year.
\end{quote}
However, not every organisation is so clearly practised or as receptive. Another participant (P1) admitted that they do not believe their current organisation has sufficient internal data governance policies and stated that additional cybersecurity and data access restrictions must be implemented first. Their interview highlighted a facet of regulatory compliance that had not been previously considered. Suppose the data structures are not adequately secured, and the access permissions are not fully established. In that case, the data is too vulnerable by default to follow the regulations' high standards. Therefore, to establish an internal policy for regulatory compliance, there needs to be some baseline of security met by an organisation in the first place. They noted having to spend a portion of their time advocating for public policy and the importance of regulatory compliance.
 
Whether an organisation has successfully implemented rules and regulations in their internal policies, there is a need for information dissemination and onboarding. Even with our current suite of technical collaboration and data storage tools, this poses a challenge.

\subsection{Knowledge Mismanaged}
Every participant reported regardless of their years of experience or the size of their organisation, a struggle with knowledge management. In this case, it is worth expanding on the current state of KM within these organisations.
\begin{quote}
    P4: I mean, we do have wikis and we do have lots of documents, but the trouble is things get out of date. So we have a huge SharePoint repository with which we've been using for, I don't know how many years, but it's probably 20 years or something like that.
\end{quote}
KM tools, like Intranets and Wikis, work well enough as document stores and ways to share information with others. Nonetheless, there remain clear challenges regarding versioning, updating, and communicating. In particular, our participants (P1 and P6) noted the speed at which Wikis fall out of date.

The current tools used by our participants fulfil only three basic parts (package, store, and transfer) of the features proposed in the KM literature \autocite{Despres1999KnowledgeS, Avram2006AtSoftware}. In other words, the current tools act more as information repositories but do not facilitate the tasks at hand to their greatest potential.
 
A potential solution for this would be to turn towards automation to process all the information and perform the functions needed by the DPOs and privacy professionals. We heard from our participants that automation is primarily unexplored territory for them. While tools like OneTrust exist, our participants noted limited experience with such tools. The primary functionality of the tool that did attract their attention was the automated Data Mapping capabilities. Our participants repeatedly mentioned that automation is either an aspiration or a challenge for their work.
\begin{quote}
    P3: If I have control of the future I may think about some sort of automation system. I don't know what it is yet, but mainly move a bit from manual work to automation.
\end{quote} 

\begin{quote}
    P7: We do not use automation in our data protection processes.
\end{quote} 
Our participants expressed the need to improve and expand the tools currently used to achieve regulatory compliance. In a world that is growing increasingly aware of the need to regulate the rapidly changing technological landscape, DPOs are working to ensure each new regulation does not require a massive overhaul for organisations.

\subsection{Findings Summarised: Knowledge Management is Key}

\begin{figure}
    \centering
    \includegraphics[width=9cm]{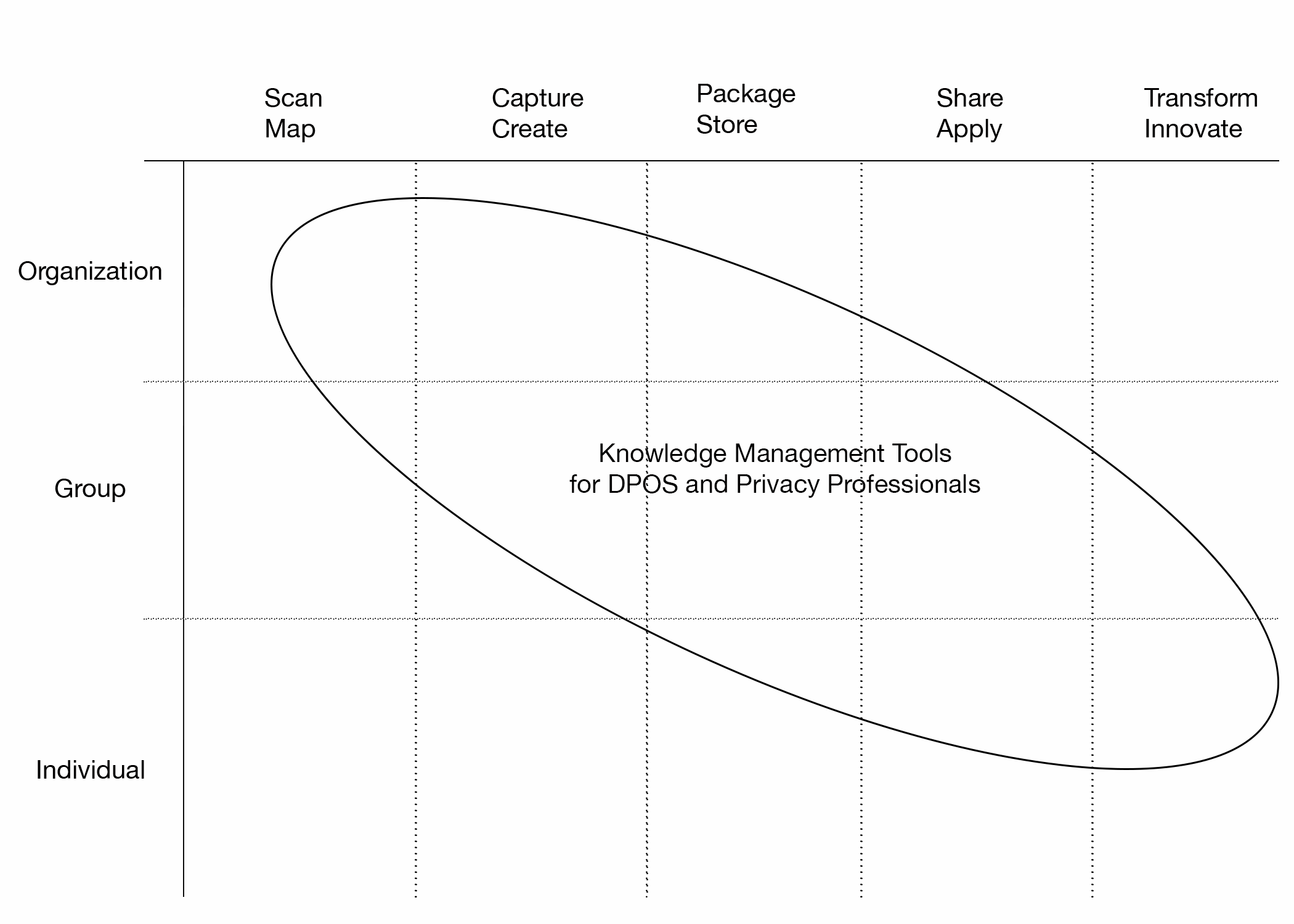}
    \caption{Inspired by Despress and Chauvel's chart mapping the regions of practice in knowledge management, this shows which regions are covered by a knowledge management tool that addresses the feedback presented by our participants \autocite{Despres1999KnowledgeS}. \protect}
    \label{fig: chart1}
\end{figure}

From our findings, the four significant categories all tie back to the idea of effective internal organisation and information processing. Each of our participants also described the role of KM as essential but also the most challenging when it comes to regulatory compliance in practice. We can see how each category of feedback corresponds with a solution offered by the KM taxonomy presented by Despres and Chauvel \autocite{Despres1999KnowledgeS}. Figure \ref{fig: chart1} shows the features, defined by Despres and Chauvel, of the KM tools for DPOs would potentially inhabit. The upper horizontal axis shows the functions a tool can play while the vertical axis identifies the level of implementation. In the first category of feedback we heard about a disconnect between regulation and practice; this may be solved by a tool able to map and capture all of the data and policies needed for an organisation to be compliant (see the upper left of Figure \ref{fig: chart1}).
 
The second category of feedback revealed how participants see their roles nebulously defined by their employers in relation to policymakers expectations. Some participants shared that the internal process of handling regulatory compliance needs to be grounded and accessible at the heart of the organisation’s workflow. In the discussion of the Bednar et al. paper, there is a division of challenges to achieving privacy engineering in practice that is labelled as the engineers’ “burden”, “inner conflict”, and “battle with lawyers” \autocite{Bednar2019EngineeringChallenge}. The interviews with our participants indicate that better KM tools will allow them to face these challenges. For example, in the related literature, it is stated that engineers often encounter a conflict between the need to ensure privacy practices while also meeting and addressing the business model of their employer \autocite{Spiekermann2019InsideEngineers}. A tool able to package information for each group or team may alleviate the tension our participants shared (see middle portion of Figure \ref{fig: chart1}).
 
In addition to the challenge of defining roles and responsibilities, the KM systems used by our participants need to process and communicate information across multidisciplinary teams. Some of the most significant challenges reported have been adjusting the language of privacy and regulatory compliance for different audiences. Once again, this is akin to similar findings from the related literature where communication between engineers and legal teams struggled \autocite{Langheinrich2002D15.3.1-AWP15-Privacy, Bednar2019EngineeringChallenge}. From previous literature and our participants’ discussions, in the third feedback category, we argue that data protection and privacy are by nature collaborative. As a result, any KM tool utilised in this space must be effective for sharing and applying regulation in an effective way for each of the different teams (see the bottom-right of Figure 
\ref{fig: chart1}).
 
Our participants and their peers must advocate for and develop internal policies that allow the business to comply with regulations. They described that establishing internal data protection and privacy policy in direct contact with the organisation’s process will provide engineers, executives, and legal personnel with a single reference source. In our discussion, we will highlight how KM literature argues for integrating this reference source into various tasks and responsibilities inherent to an organisation.

\section{Limitations}
This paper relied on an interview-based study, and the usual limitations of this kind of study apply.
%We must recognise the small sample size gathered for this paper. This limited sample automatically challenges the ability to draw large-scale generalizable conclusions. During the recruitment process, we reached out to many DPOs and other privacy professionals; only the current sample was willing to be interviewed and within that sample, only some were comfortable being quoted directly. Nonetheless, our sample size was comparable to previous similar interview studies of 6 and 12, respectively (see \autocite{Bednar2019EngineeringChallenge} and \autocite{Sirur2018AreGDPR}). Ultimately, the challenge of accessing this population of industry professionals was unexpected and worth sharing for more extensive discussion amongst data protection academics and policymakers.

We recognize that it is challenging to draw conclusions regarding compliance behaviour across different jurisdictions and legislation (USA, UK, and EU). However, it is important to note that our participants consistently need to work across all these jurisdictions regardless of their original background or training. Their organisations operate around the world yet often have only one official DPO.

\section{Discussion}
The exploratory study presented within reveals industry professionals' lived experiences and attitudes. Their statements can help inform academics and developers in establishing future tools that will effectively address the highlighted issues. Additionally, reflecting on what has been presented may provide policymakers with questions to spark essential conversations on improving data protection and privacy legislation.

\subsection{The Future of Knowledge Management for DPOs}
Concerning the challenge of KM, the issue is a matter of problem framing and terminology. The current understanding of “knowledge management” predominantly describes archiving and sharing tools that allow users to deposit, organise, and occasionally present information. However, we would argue that this does nothing to satisfy the management part of the description. Over twenty years ago, an early KM taxonomy outlined additional tasks essential for proper KM \autocite{Despres1999KnowledgeS}. The current tools we use do not prioritise information for us; they do not highlight what has gone out of date or perhaps even unusable. This lack of actual management done by the tools results in a lack of potential usability. It avoids the original goal of computation to improve human lives.

An analogy from the physical world to further visualise our understanding of KM would say the current tools marketed as knowledge management items are more like a bank vault. We can put information inside it and organise it however we wish, but once it is in there and we leave the vault, it stays the same until we return. The KM tools we aspire to are items that act more like an asset manager. Depositing the same assets as with the vault into the care of the asset manager, providing some specifications, and then allowing the system to manage the assets in such a way as will likely benefit the user when they return.

Making the jump from the vault to the asset manager has often required the intervention of a human and the specific services they can offer in addition to the basic security of physical storage space. Early work in KM also involved the need for agents and humans before the development of modern software tools entirely took over the field \autocite{Tyndale2002AApplications, Martensson2000ATool, Dingsyr2002AProject, Bjrnson2008KnowledgeTechnology}. However, with computation, we are on the cusp of facing a similar jump from large, relatively static data stores to a new future where the potential use of data is not storing it but instead letting it circulate and work for us. Let the computation handle these new tasks of processing.

Based on the information we have heard from our participants, we call for further development of KM tools to account for the lack of “innovative leverage” the tools have \autocite{Despres1999KnowledgeS}. This topic is not radical for the field of KM, but the technical tools we have built thus far have been concerned primarily with data storage and sharing. Even recent work on new KM capabilities focuses on the storage aspect \autocite{Shukri2019AInstitutions, Wang2014AEnvironment, Agarwal2014InitiatingTemplate}. While cloud storage might further alleviate the challenges faced by various organisations, the need for responsive, automated KM is clear.

Covering the whole range of possible taxonomies of a KM, from knowledge mapping to innovating and transforming, will require our tools to fill previously unexplored roles. While the possibilities for new forms of KM systems are enticing, we need to consider the impact of automation on regulatory compliance. Simply reducing the role and need for compliance to an automatic box-ticking exercise would not be valuable  \autocite{Asprion2013AssimilationInvestigation}. However, combining the automation with the work of our participants as computer-supported work may alleviate the challenges we and the literature have observed while also meeting the standards of the policymakers \autocite{Kingsto2017UsingRegulation}. This requires careful human-centred design approaches that consider the needs of all the stakeholders involved in the system. Therefore, in future work, we recommend developing knowledge management software that genuinely moves beyond a data store with sharing capabilities. 

\subsection{Reflections on the Data Protection Officer Role} \label{policy}
Compliance, cybersecurity, law, and data manifest themselves within organisations across the globe in a fine, often entangled, mesh. To ensure that we can meet the acceptable requirements in each field requires careful planning, effective communication, and well-organised information. Today's industry professionals in data privacy and protection are at the forefront of this growing challenge.

In this work, we interviewed a number of such industry experts to understand how they balance company and legal requirements. They expressed that while some things worked well, there were also apparent issues with how knowledge is managed by current software, especially concerning this new and growing challenge of regulatory compliance.

Our study found that current KM tools primarily function as document repositories and sharing tools. However, our participants noted that there is room for automation to drastically expand their current tools' functionality and capabilities as data protection professionals. This finding opens future research into how knowledge management software can be designed to meet better the complete taxonomy set forth by Despres and Chauvel \autocite{Despres1999KnowledgeS}. Moving beyond the primary role of document storage and sharing to complete computer-supported work will profoundly impact this industry and other fields.

While most of this paper's work focused on applying KM to the challenge of data protection and privacy, the findings reveal profound challenges across the industry. Multidisciplinary teams need to communicate effectively about how privacy plays into every level of an organisation. Additionally, knowing what information and data are being managed is essential for developing proper security and safety measures. Finally, we noticed a clear challenge during our exploratory study's recruitment and interview setup. For the few DPOs and privacy professionals willing to meet in the first place, we had to be precise about the steps we took to anonymise and protect the information of our participants.  It is unclear whether this caution comes from their knowledge as experts on data protection or the need to protect their organisation's practices. Nonetheless, accessing this population of industry professionals can be challenging and may pose problems for attempts to connect policymakers and professionals for future efforts and changes to legislation.

\section{Conclusion}
Accepting and adopting data protection and privacy methods remains challenging. Our participants mentioned that a portion of their daily efforts is working to convince and promote the importance of data protection and privacy within their organisation. Often these attempts at persuasion are formal presentations in front of executives and managers across all levels. Some of our participants would urge the importance of institutional pressure on adopting compliance software, similar to previous research  \autocite{Asprion2013AssimilationInvestigation}. To achieve data protection and privacy compliance, there needs to be future work exploring effective methods to drive corporate and institutional acceptance.

Ultimately this research is meant to provide industry professionals and legislators with examples and evidence of work being done within organisations to promote and implement data privacy practices. Collaboration and careful knowledge management are essential for discovering any solutions within the space. The need for cooperative discovery will likely increase as future regulation is drafted.

\section*{Acknowledgement}
This work was supported by the UKRI under PETRAS2 (EP/S035362/1) through the RETCON grant.
We would like to acknowledge the participants that took the time to speak with us and fill out our online survey. By taking time out of their days, their insights were able to provide us with the answers to our research questions. We would also like to thank the rest of our HCAI group and all the support we have had throughout this process.

\printbibliography

\appendix
\section{Appendix} \label{appendix}
\subsection{Interview Script}
\textbf{Introduction} Thank you for taking the time to meet with me for this interview today. Our conversation today is in connection with Organisational Record-Keeping for Data Regulation Compliance. You can find a summary of the project in the email sent ahead of this interview. Do you understand the aims and goals of this project?
Thank you for your interest. Just a reminder, the purpose of this interview is to better understand the types of tools you use in your daily working life and where they bring value or create challenges. We will not ask you for any personal details about you or your organisation. You can stop this interview at any time or ask for breaks as needed.  

Did you get a chance to read through and complete the informed consent document?
\begin{itemize}
    \item “Yes” - Fantastic, thank you. Did you have any questions before we continue?
    \item “No” - Please take the time to carefully read and complete it now. If there is anything you disagree with or have questions about please let me know. If there is anything within that document that makes you want to pull out from this interview, you can do so.
\end{itemize}
Before we begin, just one more data protection statement, to remind you that for any reason between now and 01/06/22 you may, without reporting the reason, request we delete the data you provide to us today.
If this sounds amicable to you, we can begin with the questions.

\paragraph{Questions - Choose the proper set depending on Interview Subject}

Intro/Background
\begin{enumerate}
    \item How would you describe your role within your organisation?
    \item How would you classify your organisation size wise?
    \begin{itemize}
        \item Big corporation
        \item An established business
        \item A small business
        \item A startup
    \end{itemize}
    \item How many projects do you work on at any one time?
    \item How many teams do you interface with regularly?
    \item Can you describe a little about those groups and their roles?
\end{enumerate}
Knowledge Management 
\begin{enumerate}
    \item Can you tell us about the tools your organisation uses to manage internal knowledge:
    \begin{itemize}
        \item What tools does your organisation use internally for keeping track of most of its knowledge?
        \item Tools for documents and files? Communication and collaboration?
        \item Project management tools? (e.g. Asana, Jira, Twilio)?
        \begin{itemize}
            \item If so, how much time per week do you spend adding and editing tasks in the software?
            \item Overall, does it help or hurt to use project management software?
        \end{itemize}
        \item Are there any other tools you have not yet mentioned that you use for knowledge management? What are they used for?
    \end{itemize}
    \item How often would you say you need to combine documents or point others to a “single source of truth”?
    \begin{itemize}
        \item Do you have one place you can rely on as the “single source of truth”?
        \item Does there need to be a “single source of truth” when interacting with legal compliance or government regulation?
    \end{itemize}
\end{enumerate}
Data Protection for Inside an Organisation 
\begin{enumerate}
    \item Describe the role compliance with data protection compliance regulations plays in your current position or role. Are you responsible for aspects of your organisation’s data protection compliance?
    \item Does your organisation have any specific tools it uses for keeping track of issues, decisions and resources relevant to data protection compliance?
    \begin{itemize}
        \item If so, what tools and how are they used? 
        \item If not, how do such issues and decisions get made and how are they preserved and organised?
        \item Who is responsible for their upkeep?
    \end{itemize}
    \item What works well with the system(s) you described above?
    \begin{itemize}
        \item Think about what features do you enjoy or regularly use… 
    \end{itemize}
    \item What is difficult with the aforementioned system?
    \begin{itemize}
        \item What frustrates you? What slows down your workflow?
    \end{itemize}
    \item  How do you organise Data Protection initiatives within Project Management Software?
    \begin{itemize}
        \item Does Data Protection (or a DPIA document) exist as its own project or does it map to multiple projects?
    \end{itemize}
    \item How does sharing compliance documents and records work?
    \begin{itemize}
        \item Where do these systems help and where do they cause issues?
        \item Are there other challenges with sharing externally from your organisation?
    \end{itemize}
    \item How do you have to handle items differently when working with legal consul or governing bodies?
    \item Does your organization have a centralized or global access control software or platform for data protection and access?
    \begin{itemize}
        \item If you don’t know/don’t have one, how do you keep track of what files you can and cannot access?
        \item Is there one team in your company that is dedicated to administering permissions? How often do you interact with them?
    \end{itemize}
    \item Do you have a clear understanding of the rules of who can access what data?
    \begin{itemize}
        \item For you personally?
        \item For others in your organization?
    \end{itemize}
    \item How often per week do you need to ask for access permissions to be granted?
\end{enumerate}
Data Protection from a DPO (Depends on Interview Subject)
\begin{enumerate}
    \item How do you conduct a Data Protection Audit?
    \item What problems do you see most organisations encounter with Data Protection?
    \item How do you support or work with multiple organisations?
    \item Can you walk us through how you handle the various types of documentation needed for the following items? 
    \begin{itemize}
        \item Article 30 Record of Processing Activities
        \item Privacy Notices
        \item Consent
        \item Access Requests
        \item Data Protection Impact Assessments
        \item Legitimate interests assessment
        \item Personal Data Breaches
        \item Data Protection Act 2018 - Special Category or Criminal Conviction and Offence data
    \end{itemize}
    \item Which of the GDPR’s required privacy policy documents is the most difficult to manage or process?
    \item What do you hope to see change in the next 5 years in regards to Data Protection Policy and Enforcement?
    \item If you had a magic wand, what would you change or alter about how you go about ensuring data protection?
\end{enumerate}
Project Management (Depends on Interview Subject) \\
If not-PMO or organizational role: “Do you work with a project manager/PMO?”
\begin{enumerate}
    \item Do all projects have an assigned project manager?
    \begin{itemize}
        \item What qualifies a project to have an assigned project manager?
        \item Personally, do you think project managers add or reduce work?
    \end{itemize}
\end{enumerate}
If PMO/Organizational Role:  
\begin{enumerate}
    \item Do you own any responsibility for assigning permissions to data used by your team?
    \begin{itemize}
        \item If yes, where do you seek guidance for allowing permissions?
        \item If no, who is the gatekeeper in your organization for data access roles?
    \end{itemize}
    \item Is there a centralized system in which you keep your project files and the data they might contain?
    \begin{itemize}
        \item Do you have any dependencies on other teams before you can access data?
        \item After use, do you store data for re-use in other projects, how so?
    \end{itemize}
    \item What do you regularly use to communicate with these teams?
\end{enumerate}
\textbf{Conclusion/Debrief} Thank you for taking the time to answer our questions, we appreciate it. The purpose of these questions is to help us establish a greater understanding of the ecosystem of tools and document managers currently in existence. We are interested in seeing how record-keeping cultures are different across teams and organisations of different sizes. We will ensure that there is no collection of any names or personally identifiable information included in our notes and final paper drafts. However, if you have any questions or concerns you can reach one of us directly at \textbf{EMAIL REDACTED FOR ANONYMITY}. There is also information within the informed consent document providing contact information to the Research Ethics Committee in case there is anything you want to discuss with them without my knowledge.

\end{document}